\documentclass[aps,amsmath,amssymb,showpacs,citeautoscript,reprint,floatfix,groupedaddress,superscriptaddress]{revtex4-1}
\usepackage[dvips]{graphicx}
\usepackage{dcolumn}
\usepackage{color}
\usepackage{bm}

\begin{document}

\title{Metastable cubic and tetragonal phases of transition metals predicted by density-functional theory}

\author{Stephan Sch\"onecker}
\email{stesch@kth.se}
\affiliation{Applied Materials Physics, Department of Materials Science and Engineering, Royal Institute of Technology, Stockholm SE-10044, Sweden}
\affiliation{IFW Dresden, P.O.\,Box 270116, D-01171 Dresden, Germany}
\author{Xiaoqing Li}
\email{xiaoqli@kth.se}
\affiliation{Applied Materials Physics, Department of Materials Science and Engineering, Royal Institute of Technology, Stockholm SE-10044, Sweden}
\author{Klaus Koepernik}
\affiliation{IFW Dresden and Dresden Center for Computational Science, P.O.\,Box 270116, D-01171 Dresden, Germany}
\author{B\"orje Johansson}
\affiliation{Applied Materials Physics, Department of Materials Science and Engineering, Royal Institute of Technology, Stockholm SE-10044, Sweden}
\author{Levente Vitos}
\affiliation{Applied Materials Physics, Department of Materials Science and Engineering, Royal Institute of Technology, Stockholm SE-10044, Sweden}
\affiliation{Department of Physics and Astronomy, Division of Materials Theory, Uppsala University, Box 516, SE-75120, Uppsala, Sweden}
\affiliation{Research Institute for Solid State Physics and Optics, Wigner Research Center for Physics, Budapest H-1525, P.O. Box 49, Hungary}
\author{Manuel Richter}
\affiliation{IFW Dresden and Dresden Center for Computational Science, P.O.\,Box 270116, D-01171 Dresden, Germany}

\begin{abstract}
By means of density-functional calculations, we systematically investigated 24
transition metals for possible metastable phases in body-centered tetragonal structure (bct), including face-centered cubic (fcc) and body-centered cubic (bcc) geometries.
A total of 36 structures not coinciding with equilibrium phases were found to minimize the total energy for the bct degrees of freedom. 
Among these, the fcc structures of Sc, Ti, Co, Y, Zr, Tc, Ru, Hf, Re, and Os, and bct Zr with $c/a=0.82$ were found to be metastable 
according to their computed phonon spectra.
Eight of these predicted phases
are not known from the respective pressure-temperature phase diagrams.
Possible ways to stabilize the predicted metastable phases are illustrated. 
\end{abstract}


\maketitle

\section{\label{sec:intro}Introduction}

A metastable configuration of an element, alloy, or compound possesses
different, perhaps superior, properties than its stable counterpart. 
The most spectacular demonstration of such behavior is probably embodied in carbon whose mechanical, thermal, and optical properties vary greatly with the allotropic form~\cite{Hirsch:2010}.

The pressure-temperature phase diagrams of 
metallic elements, 
including the technologically important
transition metals, are also rich in structural modifications~\cite{Donohue:1974,Young:1991}.
The properties of these
high-pressure or high-temperature bulk phases can, however, not be exploited under equilibrium conditions because the reverse phase transition takes place upon unloading or cooling. 
Surprisingly, the transition from a high-temperature phase to a low-temperature phase can
be suppressed in nanometer-sized materials if the particle size or grain size is smaller than a 
 specific critical value~\cite{Kitakami:1997,Rong:2006,Waitz:2009}.
An intriguing example are tens of nanometers large Co nanoparticles stabilized in the fcc phase at ambient conditions while bulk Co crystallizes in hexagonal close-packed (hcp) structure~\cite{Rong:2006b,Kitakami:1997}.
Besides cooling, other preparation routes like high-pressure torsion, milling, epitaxial growth, chemical reduction, and microwave irradiation
have been proven fruitful to stabilize pure metals in non-equilibrium crystalline phases of their phase diagram (e.g., fcc Fe nanoparticles~\cite{DelBianco:1998,Ling:2009,Kim:2003}, fcc Co films~\cite{Prinz:1985}) or to reveal \emph{unique} crystalline phases, i.e.,
phases not known from their pressure-temperature phase diagram (e.g., nanoparticles of fcc Ru~\cite{Kusada:2013}, fcc Zr~\cite{Manna:2002}, tetragonal Ag~\cite{Sun:2012}, fcc Hf~\cite{Seelam:2009}, and bcc Co films~\cite{Harp:1993}).
The common interest that drives current research on such nanostructured materials 
is related to their potentially shape- or size-enhanced mechanical, magnetic, optical, electronic, or catalytic properties~\cite{Xia:2009}.

Up to a scale of tens of nanometers, the phase stability of nano-particles is to a large extend dictated by thermodynamic surface properties, in particular by the
surface energy~\cite{Kitakami:1997,Barnard:2004,Yacaman:2001}.
The dependence of phase stability on particle-size was demonstrated for fcc Co~\cite{Kitakami:1997,Rong:2006b}, but recently theory even predicted that bulk fcc Co is
dynamically stable at 0\,K and, consequently, corresponds to a metastable bulk phase of Co~\cite{PurjaPun:2012}. 
Fcc La was found to be metastable down to very low temperatures in crystals large enough to permit the experimental determination of its phonon dispersion relations~\cite{Stassis:1985}, which were later confirmed by theory~\cite{Tutuncu:2008}.
These examples suggests a connection between non-equilibrium crystal structures observable in nanostructures and a potential isostructural metastable bulk phase.

Here, we present the results of a systematic investigation on 24 transition metal elements, that we screened
for possible metastable phases by density-functional calculations. We chose to consider the bct structure space which includes fcc and bcc structures, common to many transition metals, as special cases of higher symmetry. 
The findings of this work may encourage experimentalists to prepare metastable phases of the considered elements, e.g., as thick epitaxial films or as small particles.

The paper is structured as follows: Sec.~\ref{sec:method} elaborates on our methodological and computational approaches. 
In particular, we briefly describe the concept of the epitaxial Bain path (EBP)~\cite{Alippi:1997,Marcus:2002}, 
which we use to indentify possible metastable structures.
Criteria for elastic and dynamical stability are briefly reviewed for the sake of a self-contained presentation. Results of the computations are presented in Sec.~\ref{sec:res} and discussed in Sec.~\ref{sec:disc}, which is followed by a short summary.

\section{\label{sec:method}Methodology}

\subsection{\label{sec:method:ebp}Tetragonal minima on the epitaxial Bain path}

The bct structure with lattice parameters $a$ and $c$ coincides with the bcc (fcc) structure if $c/a = 1$ ($c/a = \sqrt{2}$).
This allows to construct homogeneous lattice transformations, referred to as Bain transformations, in the space of bct structures passing through the fcc and bcc high symmetry points~\cite{Bain:1924,Grimvall:2012}.
In general, infinitely many Bain transformations can be distinguished embodied in the freedom to define the functional relationship between $c$ and $a$. The corresponding path traversing the space spanned by $a$ and $c$ is referred to as Bain path.

The EBP is one possible realization of a Bain path, the principles of which were discussed by Alippi \emph{et al.}~\cite{Alippi:1997} and Marcus \emph{et al.}~\cite{Marcus:2002}. 
The EBP assumes those bct structures for which the lattice parameter $c_{\text{min}}$ is relaxed for every lattice parameter $a$. 
Hence, the EBP $c_{\text{min}} (a) $ is defined by the minimization of the total energy $E(a,c)$ at fixed $a$~\cite{Schoenecker:2012,Schoenecker:2015},
\begin{equation}
 E(a,c_{\text{min}}) = \min_c E(a,c).
\label{eq:ebpdef}
\end{equation}
For the total energy along the EBP we define $E_{\text{EBP}}(a) \stackrel{\text{def}}{=} E(a,c_{\text{min}} (a))$.
The EBP conditions, i.e., structural relaxation in the out-of plane direction parallel to $c$ and an isotropic distortion of the quadratic basal plane along the in-plane directions, describe the situation met in coherently grown bct films on substrates with four-fold surface symmetry.
More specifically, the EBP models the interior of the film (i.e., the bulk part) neglecting surface and interface effects except the determination of $a$ by coherency with the substrate.
For the present purpose, the EBP is a merely technical tool, as a substrate is not necessarily needed to stabilize a phase. One can imagine to grow the predicted metastable phases by epitaxy, though.

The first important symmetry-dictated demand on the total energy along the EBP arises from the fact that $E(a,c)$ is stationary at points of cubic symmetry~\cite{Kraft:1993,Craievich:1994,Marcus:2002}.
Thus, provided that the two points of cubic symmetry belong to the EBP, $E_{\text{EBP}}(a)$ is not only extremal at these two points but there is at least one additional extremal point with noncubic (bct) symmetry~\cite{Marcus:2002,Grimvall:2012}. 
In some cases, one of the cubic points does not belong to the EBP but $E_{\text{EBP}}(a)$ assumes a maximum close to this point~\cite{Schoenecker:2015}.
In the absence of spin-orbit coupling, the symmetry property also holds for ferromagnetically ordered states since the magnetization density preserves the full symmetry of the lattice.

Second, the EBP is a Bain path that passes through global or local minima of the energy $E(a, c)$ 
by virtue of its definition [Eq.~\eqref{eq:ebpdef}]. The minima of $E(a, c)$ are the minima of the EBP and referred to as tetragonal minima~\cite{Marcus:2002}. These minima possess vanishing normal stresses $\sigma_{ii}\propto \partial E/\partial \epsilon_{ii}$, 
where $\sigma_{ii}$ and $\epsilon_{ii}$, $i=\{1, 2, 3\}$, are the normal components of the stress tensor and the strain tensor, respectively.
Although these special unstressed bct states are minima of $E(a, c)$ they are not necessarily stable.
We consider bct states as stable phases, if $E_{\text{EBP}}(a)$ assumes a global minimum and the phase is known to be the experimental ground state (e.g., bcc vanadium or fcc palladium). 
Metastable phases are characterized by $E_{\text{EBP}}(a)$ at a local minimum and stability
against arbitrary small homogeneous lattice distortions and lattice vibrations. 
Unstable bct states are local minima of $E_{\text{EBP}}(a)$ which are stable only against homogeneous distortions preserving the bct symmetry. 

The present task is to identify those bct minima of $E_{\text{EBP}}(a)$ which do not coincide with an equilibrium phase and to characterize their vibrational properties. The conditions for dynamical and elastic stability are discussed in the next section.

\subsection{\label{sec:method:stab}Dynamical and elastic stability of tetragonal minima}

The criterion for dynamical lattice stability (DS) in the harmonic approximation and at zero external load 
reads~\cite{Grimvall:1999}
\begin{equation}
 \nu^2(\mathbf{q},s)>0,
\end{equation}
where $\nu$ denotes the frequency, $\mathbf{q}$ the wave vector, and the index $s$ the polarization and branch of the phonon modes. A lattice instability associated with an unstable acoustic long-wavelength phonon mode is called elastic instability. In this case, the lattice instability is often related to negative values of an elastic constant (or combinations thereof), see Table~\ref{table:christoffel:bct} for different high-symmetry branches in the bct structure.

\begin{table}
\centering
 \caption{\label{table:christoffel:bct}Relation between elastic constants and sound waves along high-symmetry branches for the bct structure obtained by solving the Christoffel equation~\cite{Grimvall:1999}. The polarization vector is given in parentheses for nontrivial cases. The four-fold rotation axis of the bct lattice is oriented along $[001]$.}
\begin{ruledtabular}
\begin{tabular}{*{4}{l}}
mode 		& $[0\xi 0]$ & $[00\xi]$ & $[\xi\xi 0]$ \\ 
\hline
longitudinal (L)   & $c_{11}$ & $c_{33}$ & $\frac{c_{11}+c_{12}+2c_{44}}{2}$ \\
transverse  (T)    & $c_{66}$ ($[100]$) & $c_{44}$ & $c_{44}$ ($[001]$)  \\
                & $c_{44}$ ($[001]$) & & $\frac{c_{11}-c_{12}}{2}$ ($[1\bar{1}0]$)  \\
\end{tabular}
\end{ruledtabular}
\end{table}

There are six independent elastic constants for the bct structure, $c_{11}$, $c_{12}$, $c_{33}$, $c_{13}$, $c_{44}$, and $c_{66}$, while cubic systems possess three, $c_{11}$ ($=c_{33}$), $c_{12}$ ($=c_{13}$), and $c_{44}$ ($=c_{66}$)~\cite{Nye:1960}.
Elastic stability in the absence of external forces is defined by a positive definite total energy for any small homogeneous deformation, which implies restrictions on the elastic constants expressed by the Born stability criteria (SCs)~\cite{Grimvall:1999,BornHuang:1954}. 
The four SCs in systems with bct symmetry are
\begin{subequations}
\begin{eqnarray}
c_{66} &>& 0 \label{eq:SC1t} \\ 
c_{44} &>& 0 \label{eq:SC2t} \\ 
c_{11}-|c_{12}| &>& 0  \label{eq:SC3t} \\ 
c^{\phantom{2}}_{33}(c^{\phantom{2}}_{11} + c^{\phantom{2}}_{12}) - 2c^2_{13}&>& 0.  \label{eq:SC4t}
\end{eqnarray}
\end{subequations}
which reduce to three SCs in the case of cubic symmetry,
\begin{subequations}
\begin{eqnarray}
c_{44} &>& 0 \label{eq:SC1c} \\ 
c_{11}-|c_{12}| &>& 0 \label{eq:SC2c} \\ 
c_{11}+2c_{12}&>&0. \label{eq:SC3c}
\end{eqnarray}
\end{subequations}
Elastic stability requires the fulfillment of all SCs.
Note that the elastic constants of the fcc structure are given with respect to the cubic axes, which are related to the elastic constants in the bct axes by a coordinate transformation, see Appendix~\ref{app:fcc}. The same coordinate transformation (Appendix~\ref{app:fcc}) needs to be applied to obtain relations between the elastic constants and sound waves in the fcc axes.

The bulk modulus ($B$) of cubic structures is related to the elastic constants by $B=(c_{11}+2c_{12})/3$, implying that SC~\eqref{eq:SC3c} is always fulfilled for cubic minima of the EBP.
In the actual calculations for tetragonal minima (Sec.~\ref{sec:res}), we find that an elastic instability never occurred due to violation of SC~\eqref{eq:SC4t}. As pointed out by Marcus \emph{et al.}, the left hand side of SC~\eqref{eq:SC4t} is related to the (positive) curvature of $E_{\text{EBP}}$ at tetragonal minima~\cite{Marcus:2002}.

\subsection{\label{sec:comp:etot}Total-energy and elastic-property calculations}

High-precision density-functional theory calculations were carried out with the full-potential local-orbital scheme FPLO-7.00-28~\cite{Koepernik:1999}.
Using the local-density approximation according to Ref.~\onlinecite{Perdew:1992} (PW92)
and a scalar-relativistic mode for elements with atomic number $< 49$, a full-relativistic mode otherwise,
we scanned the EBPs of 24 
elements with the atomic numbers 21-23, 27-30, 39-47, 72-79, in a wide range of parameters $a$~\cite{Schoenecker:2011}.
We omitted the elements Cr and Mn, which are anti-ferromagnetic with
a complicated ground state structure, and Fe, which bcc ground state is known
to be not obtained in PW92. Fe is, however, known to be elastically unstable
at the bct local minimum of its EBP, see below.

The convergence of numerical parameters and the basis set were carefully checked~\cite{Schoenecker:2011}.
In order to converge the total energy per atom at a level smaller than $0.3$\,meV, linear-tetrahedron
integrations with Bl\"ochl corrections were performed on a $24 \times 24 \times 24$ mesh in the full Brillouin zone, apart from the elements Cu, Ru, Ir, and Pt, for which a denser $48 \times 48 \times 48$ mesh was required.
To account for the possibility of ferromagnetic order in the evaluation of the EBP, total energy minimization was done with respect to both $c$ and the magnetic moment.
The bct states of the EBP were modeled by the space group $I4/mmm$. All reported  energies and moments are given per atom.

The six elastic constants of bct states were derived from fitting a square polynomial to computed total energy changes for small strain deformations applied to the bct reference cell (tetragonal axes) following Ref.~\onlinecite{Jona:2001}.
The elastic constants $c_{11}$, $c_{12}$, and $c_{13}$ were modeled by space group $I/mmm$, $c_{33}$ by space group $I4/mmm$, $c_{44}$ by space group $C2/m$, and $c_{66}$ by space group $F/mmm$.
The largest imposed moduli of the strains were $0.5\,\%$ ($c_{11}$, $c_{12}$, $c_{33}$, $c_{13}$, and $c_{66}$) and $0.75^\circ$ ($c_{44}$). All elastic constants were found to be stable within $5\,\%$ change against a doubling of the strain interval. The doubling of the strain interval did not affect the conclusion on elastic stability.

All here reported elastic constants of the fcc structure and of the bct (bcc) structure are given with respect to the cubic axes and with respect to the bct (bcc) axes, respectively. 
We verified numerically that deformations applied to both coordinate systems yield virtually identical elastic constants for all investigated fcc structures.

\subsection{\label{sec:comp:phon}Phonon calculations}

The force-constant matrix was obtained within the framework of density-functional perturbation theory~\cite{Baroni:2001} as implemented in the Vienna \emph{ab initio} simulation package~\cite{Kresse:1996} based on the in the projector-augmented wave formalism~\cite{Bloechl:1994,Kresse:1999}. The software ``Phonopy''~\cite{phonopy}  was employed to determine the phonon dispersion relation from the force-constant matrix and the phonon density of states (DOS) by linear-tetrahedron integration.

Convergence of all numerical parameters was carefully checked. We used a plane wave cut-off of $400$\,eV for the elements Sc and Y, and a larger cut-off ($500$\,eV) otherwise.
Brillouin zone integrations were performed on a $k$-point mesh equivalent to a $36\times 36\times 36$ Monkhorst-Pack mesh of the primitive unit cell smeared by a first order Methfessel-Paxton scheme~\cite{Methfessel:1989b} with small smearing parameter (0.1\,eV).
The lattice parameters of the bct states (cubic states) in question were relaxed until the residual stress along the tetragonal axes (cubic axes) was smaller than $\sim 3$\,bar prior to computing the force-constant matrix. The phonon properties were sampled on a $4\times 4\times 4$ $\Gamma$-centered mesh and convergence was checked against a denser $6\times 6\times 6$ $\mathbf{q}$-mesh. In the singular case of bct Zr, the computation of the phonon properties required the denser $\mathbf{q}$-mesh.

To verify the accuracy of our approach, we first reproduced the experimental phonon dispersion curves and DOSs for fcc Au and bcc Fe for which reliable experimental data are available (results not shown)~\cite{Schober:1981}.

\section{\label{sec:res}Results}

Elements with stable hcp ground state structure and elements with stable cubic ground state structure are discussed separately in the following to appropriately reflect the similarities found in the structure of their EBPs and in their stability behaviors.
The elastic constants and SCs for the determined cubic and tetragonal minima of the EBP were investigated in all cases, but dynamical stability was analyzed only for elastically stable structures.

\subsection{\label{sec:reshcp}Elements with stable hcp structure}

\begin{table}[tb]
\centering
\caption{\label{table:hcptable}Stability analysis of minima on the EBP for hcp stable elements (first column, sorted by atomic number). The next four columns classify the symmetry (sym) of the minimum (C is cubic, N is noncubic), if ferromagnetic (fm) order is present, the lattice parameter ($a$), the lattice parameter ratio ($c/a$), and the energy difference with respect to the ground state ($\Delta E$). Columns 6 - 8 and columns 9 - 10 give the signs of the right hand side of the tetragonal SCs~\eqref{eq:SC1t} -~\eqref{eq:SC3t} and the cubic SCs~\eqref{eq:SC1c} and~\eqref{eq:SC2c}, respectively. SC~\eqref{eq:SC4t} or~\eqref{eq:SC3c} is fulfilled for all minima and omitted. If the signs of all SCs are positive, the last column indicates if the minimum is dynamically stable at 0\,K ($\boldsymbol{\times}$ stable, $-$ unstable).}
\begin{ruledtabular}
\begin{tabular}{ll*{2}{c}@{}r*{4}{c@{}}cc}
element & sym & $a$ & $c/a$ & \multicolumn{1}{c}{$\Delta E$} & \multicolumn{5}{c}{SC} & DS \\
\cline{6-10}
 & & (\textrm{\AA}) & & \multicolumn{1}{c}{(meV)} & \eqref{eq:SC1t} & \eqref{eq:SC2t} & \eqref{eq:SC3t} & \eqref{eq:SC1c} & \eqref{eq:SC2c}\\
\hline
Sc & C & 3.17 & $\sqrt{2}$ & 35 &  &  &  & + & + & $\boldsymbol{\times}$ \\
   & N & 3.70 & 0.90 & 116 & + & + & $-$ &  & & \\
Ti & C & $\sqrt{2}$ & 54 &  &  &  & + & + &  $\boldsymbol{\times}$ \\ 
   & N & 3.33 & 0.86 & 72 & + & + & + &  & & $-$ \\
Co & C (fm) & 2.42 & $\sqrt{2}$ & 25 &  &  &  & + & + & $\boldsymbol{\times}$ \\
   & N (fm) & 2.81 & 0.92 & 137 & + & + & $-$ & & & \\ 
Zn & C & 2.68 & $\sqrt{2}$ & 27 &  &  &  & $-$ & + & \\ 
   & N
 & 3.30 & 0.77 & 29 & + & + & $-$ &  &  &  \\ 
Y & C & 3.47 & $\sqrt{2}$ & 10 &  &  &  & + & + & $\boldsymbol{\times}$ \\
  & N & 4.07 & 0.88 & 113 & + & + & $-$ & &  & \\
Zr & C & 3.14 & $\sqrt{2}$ & 33 &  &  &  & + & + & $\boldsymbol{\times}$ \\   
   & C & 3.49 & 1 & 47 &  &  &  & + & + & $-$ \\ 
   & N
& 3.74 & 0.82 & 32 & + & + & + &  & &  $\boldsymbol{\times}$ \\
Tc & C & 2.71 & $\sqrt{2}$ & 87 &  &  &  & + & + & $\boldsymbol{\times}$ \\
   & N & 3.14 & 0.90 & 287 & + & + & $-$ & &  &\\
Ru & C & 2.66 & $\sqrt{2}$ & 125 &  &  &  & + & + & $\boldsymbol{\times}$ \\
   & N (fm) & 3.18 & 0.85 & 598 & + & + & $-$ & & & \\
Hf & C & 3.09 & $\sqrt{2}$ & 62 &  &  &  & + & + & $\boldsymbol{\times}$ \\
   & N & 3.64 & 0.85 & 115 & + & + & + &  & & $-$ \\
Re & C & 2.74 & $\sqrt{2}$ & 95 &  &  &  & + & + & $\boldsymbol{\times}$ \\
   & N & 3.18 & 0.90 & 381 & + & + & $-$ &  & & \\
Os & C & 2.70 & $\sqrt{2}$ & 136 &  &  &  & + & + & $\boldsymbol{\times}$ \\
   & N & 3.24 & 0.84 & 751 & + & + & $-$ & & &\\
\end{tabular}
\end{ruledtabular}
\end{table}

The analysis of the EBPs for eleven investigated elements with hexagonal ground state revealed that all but one possess two minima on the EBP, see Table~\ref{table:hcptable}. The notable exception is Zr, which possesses one additional minimum~\cite{Ji:2003}. 
Common to all investigated elements except Zr is that the fcc structure ($c/a = \sqrt{2}$) coincides with the global minimum of $E_{\text{EBP}}(a)$ and that the second, bct minimum has $c/a<1$. It follows from the general arguments provided in Sec.~\ref{sec:method:ebp} that bcc is a saddle point of the total energy  $E(a,c)$. 
In the case of Zr, the bct minimum with $c/a=0.82$ is marginally more stable than the fcc structure and the bcc structure coincides with the third minimum of $E_{\text{EBP}}(a)$, Table~\ref{table:hcptable}.

Both the hcp phase and the fcc phase (stable above 696\,K) of Co are known to order ferromagnetically which is also the magnetic ground state reproduced by the present calculations. 
We find that the bct minimum of Co exhibits ferromagnetic order with a spin magnetic moment equal to 1.63\,$\mu_\text{B}$, which is slightly larger than the calculated spin magnetic moments for hcp Co (1.51\,$\mu_\text{B}$) and for fcc Co (1.54\,$\mu_\text{B}$) and the corresponding experimental values, 1.52\,$\mu_\text{B}$ for hcp Co~\cite{Bonnenberg:1986} and 1.51\,$\mu_\text{B}$ for quenched fcc Co~\cite{Reck:1969}. The pronounced insensitivity of the magnetic moment of Co on details of the structure expresses its strong ferromagnetic character.
Bulk hcp Ru does not order magnetically, but the bct minimum of Ru is ferromagnetic as shown by some of us recently~\cite{Schoenecker:2012}.

The computed elastic constants allow the examination of the elastic SCs for all identified minima, see Table~\ref{table:hcptable}.
We find that the fcc structures of all investigated elements except Zn and bcc Zr are elastically stable, while most bct minima are unstable except those of the fourth group elements,
bct Ti ($c/a=0.86$), bct Zr ($c/a=0.82$), and bct Hf ($c/a=0.85$).
The elastic stability of the fcc structures of Sc, Ti, Co, Y, Zr, Tc, Ru, Hf, Re, and Os~\cite{Aguayo:2002,Ji:2003,Wills:1992,Papaconstantopoulos:2001,Marcus:1997,Wang:2003,Guo:2000}, the elastic stability of bct Zr~\cite{Ji:2003} as well as the elastic instability of bct Co, fcc and bct Zn~\cite{Marcus:2002} was reported previously in accordance with the findings of the present work.
There is, however, a discrepancy on the elastic stability of bcc Zr and bct Ti. 
Bcc Zr was was found to be stable in Ref.~\cite{Ji:2003} and in this work and was reported to be unstable in Refs.~\cite{Porta:2001,Ikehata:2004}.
The point at issue is related to the long-wavelength part of the transverse phonon branch along $[\xi \xi 0]$ polarized along $[1\bar{1}0]$, $T_{[1\bar{1}0]}[\xi\xi0]$, i.e., the sign of the shear constant $C'\equiv (c_{11}-c_{12})/2$ (Table~\ref{table:christoffel:bct}). 
An important difference in the two sets of results is that the former two works were obtained using the local-density approximation to exchange and correlation while the latter two employed a gradient-corrected functional (PBE96)~\cite{Perdew:1996,*Perdew:1996E}.
We computed $C'$ of bcc Zr choosing PBE96 and indeed obtained a negative value. 
Bct Ti is a similar case as Marcus \emph{et al.}~\cite{Marcus:2002} reported $C'<0$ using PBE96, which we were able to reproduce, but PW92 gives $C'>0$.
Thus, elastic stability of bcc Zr and bct Ti depends on the chosen exchange-correlation functional. All three unclear cases turn out to be dynamically
unstable. Thus, the undecided elastic instability does not pose any practical
issue.

\begin{figure*}[t]
\resizebox{2\columnwidth}{!}{\includegraphics[clip]{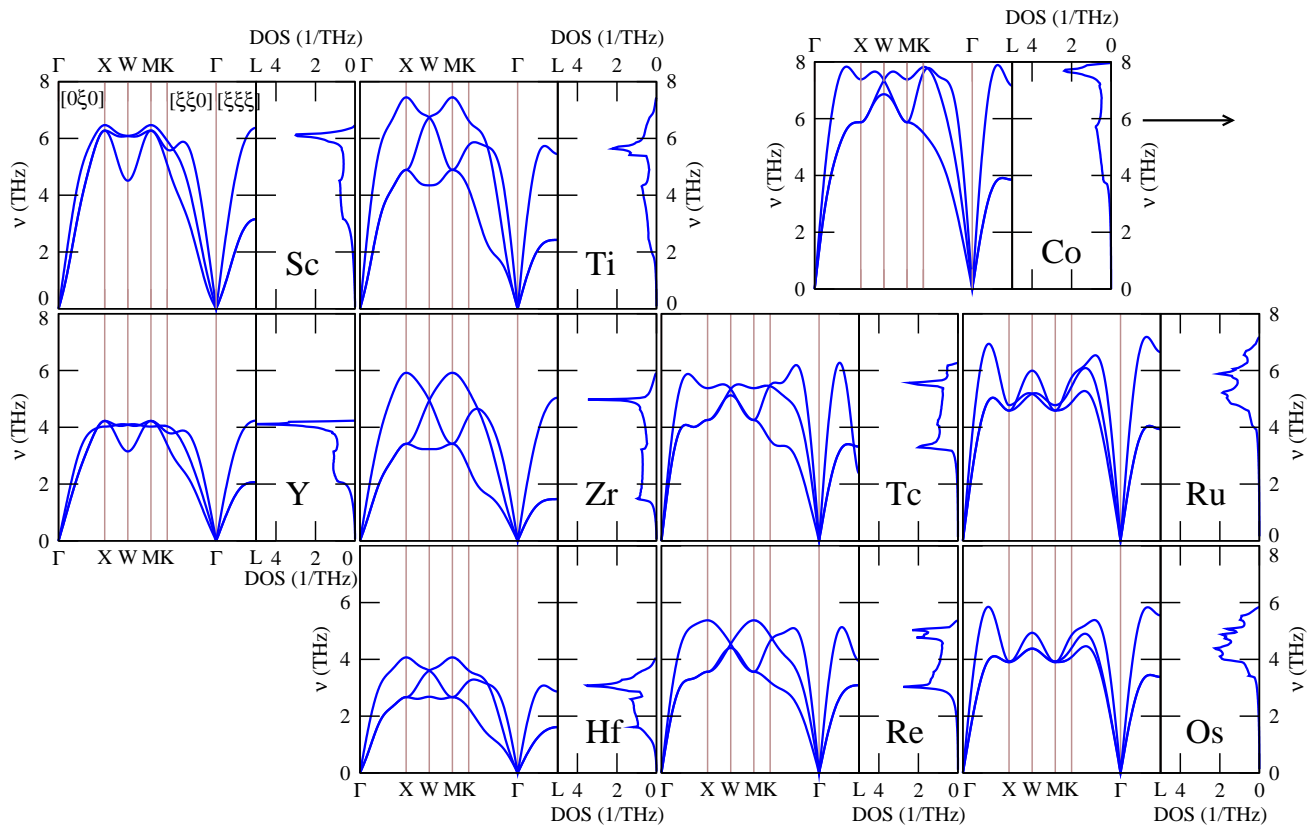}}
\caption{\label{fig:phonhcp}(Color online)
Phonon dispersion curves and DOSs for the elastically stable fcc structures of hcp stable elements according to Table~\ref{table:hcptable}. Note that Co should be in the rightmost position according to its group in the periodic system of elements (indicated by an arrow). 
The DOSs are normalized to the number of normal modes per atom (3). The high-symmetry branches denoted only for the dispersion curves of Sc apply to all panels.}
\end{figure*}

Elastic stability is a necessary, but not a sufficient criterion for structural (meta)stability of a phase. The phonon dispersion curves and derived phonon DOS allow to investigate dynamical stability as a sufficient criterion.
Both phononic properties are plotted in Fig.~\ref{fig:phonhcp} for the ten elastically stable fcc structures revealing that all structures (Sc, Ti, Co, Y, Zr, Tc, Ru, Hf, Re, and Os) are in fact \emph{dynamically stable}. We note that dynamical stability of fcc Co and fcc Re at 0\,K was reported in previous studies of their lattice dynamical properties~\cite{Persson:1999,PurjaPun:2012}.

\begin{figure}
\resizebox{\columnwidth}{!}{\includegraphics[clip]{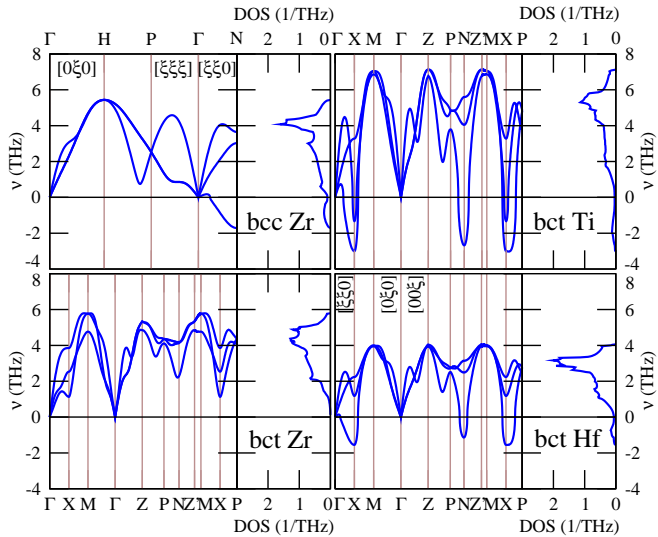}}
 \caption{\label{fig:phonbctbcc}(Color online) Phonon dispersion curves and DOSs for the elastically stable bcc and bct structures for elements with stable hcp ground state according to Table~\ref{table:hcptable}. The DOS are normalized to the number of normal modes per atom (3). -$|\nu|$ is plotted when $\nu^2<0$. The high-symmetry branches for bct structures denoted only for the dispersion curves of Hf also apply to bct Ti and Zr.}
\end{figure}

Among the four elastically stable structures bct Ti, Zr, Hf, and bcc Zr,
only bct Zr is found to be dynamically stable, see Fig.~\ref{fig:phonbctbcc} for the corresponding phonon dispersion curves and DOSs. Bcc Zr has imaginary phonon frequencies around the $N$ point. Bct Ti and Hf exhibit imaginary phonon frequencies around the $X$ and $N$ points. These two zone-boundary phonons are softened in bct Zr which may indicate marginal stability of this phase.

The elastic parameters and Debye temperatures of all metastable fcc phases (Sc, Ti, Co, Y, Zr, Tc, Ru, Hf, Re, and Os) and bct Zr are tabulated for completeness in Appendix~\ref{app:elasconstfcc}. The knowledge of elastic constants is essential, e.g., for predicting the phase stability of fine particles, and is required as input for higher order linear-elasticity continuum approaches and phase-field simulations~\cite{Kitakami:1997,Kundin:2010}.

\subsection{\label{sec:rescub}Elements with stable cubic structure}

Focusing first on the investigated elements with stable fcc phase, we find that all EBPs possess two minima in total energy. The global minimum coincides with the fcc structure and the second minimum is bct with $c/a <1$, see Table~\ref{table:fccstable}. We note that the $c/a$ ratio of the second minimum increases as the transition metal series are traversed from left (low $d$-band occupation) to right (high $d$-band occupation).
Ferromagnetic order is present for the bct state of Ni with a spin magnetic moment equal to 0.55\,$\mu_{\text{B}}$ which is only slightly larger than the experimental spin magnetic moment of fcc Ni, 0.52\,$\mu_{\text{B}}$~\cite{Reck:1969}.

All bct minima are found to be elastically unstable with respect to the same orthorhombic deformation of the bct cell caused by a violation of SC~\eqref{eq:SC3t}, see Table~\ref{table:fccstable}, corresponding to an unstable mode $T_{[1\bar{1}0]}[\xi \xi 0]$ (Table~\ref{table:christoffel:bct}).
This elastic instability was reported previously for all investigated elements except Ni and Pt~\cite{Jona:2001,Jona:2002,Mehl:2004}.

\begin{table}[tbh]
\centering
\caption{\label{table:fccstable}Stability analysis of local minima on the EBP for fcc stable elements. Notation of columns 1 - 8 as in Table~\ref{table:hcptable}. SC~\eqref{eq:SC4t} is fulfilled for all minima and omitted.}
\begin{ruledtabular}
\begin{tabular}{ll*{2}{c}@{}r*{2}{@{}c}@{}c}
element & sym & $a$ & $c/a$ & \multicolumn{1}{c}{$\Delta E$} & \multicolumn{3}{c}{SC} \\
\cline{6-8}
 & & (\textrm{\AA}) & & \multicolumn{1}{c}{(meV)} & \eqref{eq:SC1t} & \eqref{eq:SC2t} & \eqref{eq:SC3t} \\
\hline
Ni & N (fm) & 2.86 & 0.87 & 87 & + & + & $-$ \\ 
Cu & N & 2.85 & 0.95 & 38 & + & + & $-$\\
Rh & N & 3.22 & 0.82 & 252 & + & + & $-$ \\
Pd & N & 3.20 & 0.88 & 59 & + & + & $-$\\
Ag & N & 3.31 & 0.90 & 36 & + & + & $-$  \\
Ir & N & 3.30 & 0.80 & 378 & + & + & $-$\\
Pt & N & 3.34 & 0.81 & 40 & + & + & $-$ \\
Au & N & 3.39 & 0.86 & 25 & + & + & $-$ \\
\end{tabular}
\end{ruledtabular}
\end{table}

The investigated elements with stable bcc phase show two common features of their EBPs, a global total energy minimum at the bcc structure and a second minimum with 
$c/a > \sqrt{2}$, see Table~\ref{table:bccstable}. In contrast to the fcc stable elements, the $c/a$ ratio of the second minimum decreases as the transition metal series are traversed from left to right (with increasing $d$-band occupation).
All bct minima are elastically unstable due to a negative $c_{66}$ [violation of SC~\eqref{eq:SC1t}]. A negative $c_{66}$ corresponds to an unstable long-wavelength part of the transverse phonon mode $T_{[100]}[0\xi 0]$.
Apart from Ta, this elastic instability was reported before~\cite{Marcus:2002,Mehl:2004}.
To complete the picture of the bcc transition metals, we note that Fe was predicted to order antiferromagnetically and to be unstable due to $c_{66}<0$ at the bct local minimum of its EBP~\cite{Qiu:2001,*Qiu:2001:E}.

The lattice instability related to $c_{66}<0$ corresponds to a monoclinic lattice distortion of the bct unit cell. 
The highest symmetry of the distorted lattice is, however, face-centered orthorhombic. The orthorhombic axes are related to the bct reference frame through a rotation around $[001]$ by $\pi/4$. 
In the rotated, orthorhombic frame (primed notation), $c_{66}$ equals $(c^{\prime}_{11} - c^{\prime}_{12})/2$ (cf.~Appendix~\ref{app:fcc}).

\begin{table}[tbh]
\centering
\caption{\label{table:bccstable}Stability analysis of local minima on the EBP for bcc stable elements. Notation of columns 1 - 8 as in Table~\ref{table:hcptable}. SC~\eqref{eq:SC4t} is fulfilled for all minima and omitted.}
\begin{ruledtabular}
\begin{tabular}{ll*{2}{c}@{}r*{1}{@{}c}*{2}{@{}c}}
element & sym & $a$ & $c/a$ & \multicolumn{1}{c}{$\Delta E$} & \multicolumn{3}{c}{SC} \\
\cline{6-8}
 & & (\textrm{\AA}) & & \multicolumn{1}{c}{(meV)} & \eqref{eq:SC1t} & \eqref{eq:SC2t} & \eqref{eq:SC3t} \\
\hline
V & N & 2.40 & 1.84 & 111 & $-$ & + & + \\
Nb & N & 2.70 & 1.79 & 183 & $-$ & + & + \\
Mo & N & 2.61 & 1.75 & 382 & $-$  & + & + \\
Ta & N & 2.75 & 1.71 & 196 & $-$ & + & + \\
W & N & 2.68 & 1.66 & 403 & $-$  & + & + \\
\end{tabular}
\end{ruledtabular}
\end{table}

\section{\label{sec:disc}Discussion}

The results of the previous section indicate clear similarities among the transition metals elements crystallizing either in the hcp structure or in a cubic (fcc, bcc) structure.
The EBPs of all investigated elements but Zr possess two tetragonal minima each irrespective of their ground state phase. 
Apart from Zr, only the $5f$-element uranium is known to exhibit more than two minima on the  EBP~\cite{Schoenecker:2012}. This corroborates the earlier
finding that an EBP with two minima is indeed the \emph{standard} case for most metallic elements~\cite{Marcus:2002}. 

Considering the general structure of their EBPs, all elements with stable hcp phase, except Zr, and all elements with stable fcc phase show two common characteristics: (i) their standard EBPs exhibit the global energy minimum at the fcc structure and (ii) the local minimum is bct with $c/a<1$. 
On the other hand, the local bct minimum of bcc stable elements has an axial ratio $c/a>\sqrt{2}$, while the global minimum coincides with the bcc phase. 
These findings imply that none of the investigated elements with standard EBP has both minima at cubic structures. 
It is worth noting that there exist exceptions to this rule in the alkali and alkaline metal series~\cite{Marcus:2002}.

The metastable fcc structures of the investigated hcp elements 
lie very close in energy to the respective ground states, see Table~\ref{table:hcptable}.
Much higher energies are found for the related bct minima of the EBP or for
the bcc structures, with Zr and Zn as the only exceptions.
This behavior is universal for all non-magnetic transition metals with $d$-band occupations
$\approx$ 1.5 - 2.5 electrons (Sc and Ti groups) or $\approx$ 5.5 - 6.5 electrons (Tc, Re, and Ru, Os)~\cite{Skriver:1985} as well as for Co which has an hcp ground
state due to magnetism~\cite{Soederlind:1994c}.
The low energy of the metastable fcc structures can be simply understood from 
the same packing density of fcc and hcp. It is certainly advantageous for the preparation of thick epitaxial films.

None of the minima with bct symmetry except bct Zr with $c/a=0.82$ was found to be dynamically stable, i.e., these bct minima do not correspond to metastable bct bulk phases.
In most cases, structural instability is already signaled by a single negative elastic constant or a violated stability condition composed of a combination of elastic constants.
All elastic instabilities at bct minima of the investigated hcp and fcc stable elements arises in all cases due to a negative shear elastic constant $C^{\prime}$, while elastic instability of the bct minima of all investigated bcc stable elements consistently occurs due to a negative $c_{66}$. 
Both instabilities embody a shear of the quadratic basal plane [$(001)$ plane] of the bct structure. 
Although our findings show that these bct structures are unstable in bulk form, an appropriate external constraint may impede the shear deformation related to the instability from happening, at least on a nanometer length scale. A possible realization of such a constraint is a substrate with four-fold rotation symmetry on which the bct structure is coherently grown assuming the epitaxial relationship substrate (001) $\parallel$ film (001). Examples of coherent heteroepitaxial growth of metallic overlayers in bct structures are given in Table~\ref{table:epitaxy}.

\begin{table}
\caption{\label{table:epitaxy}Coherent heteroepitaxial growth of metallic overlayers in bct structures. The $(001)$-oriented substrate material is specified, and the measured thickness [in monolayers (ML)] of the coherent bct film and its $c/a$ ratio are denoted.
In each case, coherency of the film with the substrate was verified by low-energy electron diffraction.}
\begin{ruledtabular}
\begin{tabular}{ccccc}
substrate & film & thickness (ML) & $c/a$ & Ref. \\
\hline
Al & Ti & 12 & 1.495  & \cite{SKKim:1996c} \\
W & Zr & 17 & 1.48  & \cite{Ji:2003} \\
W & Zr & 50 & 1.46  & \cite{GEHill:1971} \\
Ni & Co & 30 & 1.45  & \cite{Chambers:1987} \\
Cu & Co & 10 & 1.36  & \cite{Cerda:1993} \\
Pd & Co & 30 & 1.13  & \cite{Giordano:1996} \\
Pt & Co & 12 & 1.07  & \cite{Valvidares:2004} \\
Pd & Ni & 12 & 1.11  & \cite{Petukhov:2003} \\
GaAs & Ni & 25 & 1.00  & \cite{Tang:2002,*Tian:2005} \\
W & Pd & 13 & 0.92  & \cite{Ji:2002} \\
Pd & Cu & 10 & 1.18  & \cite{HLi:1989} \\
Pt & Cu & 15 & 1.17  & \cite{YSLi:1991} \\
\end{tabular}
\end{ruledtabular}
\end{table}

The predictions on metastability for Sc, Ti, Co, Y, Zr, Tc, Ru, Hf, Re, and Os made in this work call for experimental verification. In the following we briefly address possible preparation routes. 
Under the circumstance that there exists a high-temperature fcc (or, bct) phase, 
appropriately cooling this phase may impede the allotropic structural transition to the stable hcp phase, and the fcc (or, bct) structure may coexist as metastable form at ambient temperatures as realized for example for fcc La~\cite{Stassis:1985}.
With the exceptions of fcc Co and Y, an fcc (or, bct) phase is, however, not observed in the pressure-temperature phase diagrams of the investigated elements~\cite{Donohue:1974,Young:1991}.
Of the two, only fcc Co is stable at high temperatures, while fcc Y was stabilized under high pressure.
A combination of pressure and shear stress has proven to induce a phase transition and to stabilize non-equilibrium phases at ambient conditions, e.g., for bcc Zr~\cite{PerezPrado:2009}. As a similar route to metastability, materials can undergo a phase transition during indentation~\cite{Mukhopadhyay:2013}.
Possible other approaches to stabilize metastable structures involve epitaxial growth or the fabrication of nanoparticles. As pointed out in Sec.~\ref{sec:intro}, the fcc structures of Zr, Co, Fe, and recently Ru and Hf were already discovered in synthesized nanometer-sized particles or films at ambient conditions. 
This experimental evidence supports the present finding on the metastability of a corresponding isostructural bulk phase.

\section{SUMMARY}

We screened, by means of density-functional calculations, the bct epitaxial Bain paths of 24 metallic elements and found 11 metastable phases, see Table~\ref{table:summ}.
Elastic stability of these phases had been predicted in earlier publications, but dynamical stability had previously been verified for only two of these phases.
On the experimental side, six phases are either known from the pressure-temperature phase diagram or were stabilized in nanoscale particles or films.
Experimental evidence is still lacking for five of the predicted phases: fcc Sc, Tc, Re, Os, and bct Zr.

\begin{table}
\caption{\label{table:summ}
Overview over the predicted metastable phases (left column). References on previous theoretical predictions (at zero pressure and temperature) of elastic or dynamical stability (ES and DS, respectively) and available experimental information (exp.) are given in the inner columns. 
The last column shows predicted lattice parameters (PL), obtained by multiplication of the lattice parameters evaluated in PW92 with $(V^{\rm hcp}_{\rm exp}/V^{\rm hcp}_{\rm PW92})^{1/3}$. Here, $V^{\rm hcp}_{\rm exp}$ and $V^{\rm hcp}_{\rm PW92}$ denote the experimental and PW92 volume
in the hcp structure, respectively. The predicted lattice parameters refer to the cubic axes for fcc phases and denote $a_{\text{bct}}$ for the bct structure. Experimental lattice parameters were taken from Ref.~\cite{Villars:1991}.}
\begin{ruledtabular}
\begin{tabular}{@{}lrrrl@{}}
phase & \multicolumn{1}{c}{ES} & \multicolumn{1}{c}{DS} & \multicolumn{1}{c}{exp.} & \multicolumn{1}{c}{PL} \\
\hline
fcc Sc & \cite{Papaconstantopoulos:2001} & - & - & 4.628 \\
fcc Ti & \cite{Papaconstantopoulos:2001,Wang:2003,Marcus:2002,Aguayo:2002} & - & \cite{Manna:2003} & 4.116 \\
fcc Co & \cite{Guo:2000} & \cite{PurjaPun:2012} & \cite{Donohue:1974,Young:1991,Harp:1993,Huang:1996,meng:2013} & 3.541 \\
fcc Y  & \cite{Papaconstantopoulos:2001} & - & \cite{Donohue:1974,Young:1991} & 5.071\\
fcc Zr & \cite{Papaconstantopoulos:2001,Ji:2003,Marcus:2002} & - & \cite{Manna:2002} & 4.517 \\
bct Zr & \cite{Ji:2003,Aguayo:2002} & - & - & 3.815\footnote{$c/a=0.82$} \\
fcc Tc & \cite{Papaconstantopoulos:2001} & - & - & 3.863 \\
fcc Ru & \cite{Papaconstantopoulos:2001} & - & \cite{Kusada:2013} & 3.793 \\
fcc Hf & \cite{Wills:1992,Papaconstantopoulos:2001,Aguayo:2002} & - & \cite{Seelam:2009} & 4.471\\
fcc Re & \cite{Wills:1992,Papaconstantopoulos:2001} & \cite{Persson:1999} &-  & 3.895\\
fcc Os & \cite{Wills:1992,Papaconstantopoulos:2001} & - & - & 3.828\\
\end{tabular}
\end{ruledtabular}
\end{table}

\section*{ACKNOWLEDGEMENTS}
The Swedish National Infrastructure for Computing at the National Supercomputer Centers in Link\"oping and Stockholm are gratefully acknowledged for providing computational facilities.
Financial support by the Swedish Research Council, the Swedish Foundation for Strategic Research, and the Hungarian Scientific Research Fund (OTKA 84078 and 109570) is acknowledged.
M.R.~likes to thank M. L\"oser and P. Paufler for discussions.

\appendix
\section{\label{app:fcc}Elastic constants of the fcc structure}

The elastic constants of the fcc structure given with respect to the cubic axes are related to the elastic constants in the bct axes through a rotation around the tetragonal axis ($[001]$) by $\pi/4$. The components of the elasticity tensor with respect to the bct axes are denoted by $c_{ijkl}$ and with respect to the cubic axes by $c^{\prime}_{ijkl}$. Both are related through tensor transformations rules~\cite{Sadd:2005} (Einstein summation convention),
\begin{equation}
 c^{\prime}_{ijkl} = L_{im}L_{jn}L_{kq}L_{lp}c^{\phantom{\text{BCT}}}_{mnqp}, 
\label{eq:tensortransformc}
\end{equation}
where the matrix $[L]$ describing the present rotation is 
\begin{equation}
 [L] = \frac{1}{\sqrt{2}}\left( \begin{array}{rrr} 1 & -1 & 0 \\ 1 & 1 & 0 \\ 0 & 0 & \sqrt{2}  \end{array}\right). \nonumber
\end{equation}
Performing the summations in Eqs.~\eqref{eq:tensortransformc}, one derives for the six independent elastic constants of the bct structure in the rotated coordinate system (Voigt notation applied~\footnote{Voigt notation: the two index pairs in matrix notation $ij$, $i,j=\{x,y,z\}$, are replaced by a single index each: $xx \to 1$; $yy \to 2$; $xx \to 3$; $yz, zy \to 4$; $xz, zx \to 5$; $xy, yx \to 6$.}),
\begin{eqnarray}
 c^{\prime}_{11} &=& \frac{1}{2} \left(c^{\phantom{\prime}}_{11} + c^{\phantom{\prime}}_{12} +2c^{\phantom{\prime}}_{66}\right)  \nonumber \\
 c^{\prime}_{12} &=& \frac{1}{2} \left( c^{\phantom{\prime}}_{11} + c^{\phantom{\prime}}_{12} -2c^{\phantom{\prime}}_{66} \right)  \nonumber \\
 c^{\prime}_{66} &=& \frac{1}{2}\left( c^{\phantom{\prime}}_{11} - c^{\phantom{\prime}}_{12} \right)  \nonumber \\ 
 c^{\prime}_{33} &=& c^{\phantom{\prime}}_{33} \nonumber \\
 c^{\prime}_{13} &=& c^{\phantom{\prime}}_{13} \nonumber \\  
 c^{\prime}_{44} &=& c^{\phantom{\prime}}_{44} \nonumber, 
\end{eqnarray}
with the usual symmetries of $c^{\prime}_{ijkl}$ for cubic structures.

\section{\label{app:elasconstfcc}Elastic constants and Debye temperatures of metastable fcc phases and bct Zr}

The elastic parameters of the metastable fcc phases (see Table~\ref{table:hcptable} in Sec.~\ref{sec:reshcp}) are listed in Table~\ref{table:app:elconst:fcc}.
The corresponding Debye temperatures, derived on the one hand from the elastic constants ($\theta^{\text{e}}_{\text{D}}$) and on the other hand from a fit of a harmonic function to the low-frequency part of the phonon DOS ($\theta^{\text{p}}_{\text{D}}$), are also tabulated.
For the computation of $\theta^{\text{e}}_{\text{D}}$, we closely followed the formalism and procedure elaborated in Ref.~\cite{Li:2012}. In order to obtain $\theta^{\text{p}}_{\text{D}}$, the DOS was fitted from $0$ to $1/7$ of the maximum phonon frequencies in the cases of Ti, Y, Zr, and Hf, and from 0 to $1/4$ of the maximum phonon frequencies in all other cases. The two choices for the cut-off ensured a fit to a  Debye-like behavior ($\propto \nu^2$) of the low-frequency part of the DOS.

The elastic parameters of metastable bct Zr (see Table~\ref{table:hcptable} in Sec.~\ref{sec:reshcp}, PW92 lattice parameter $a=3.74\,\textrm{\AA}$, $c/a=0.82$) are (in units of GPa) $c_{11}=130$, $c_{12}=92$, $c_{66}=24$, $c_{33}=134$, $c_{13}=78$, and $c_{44}=40$, and $\theta^{\text{p}}_{\text{D}}=274$\,K ($\theta^{\text{e}}_{\text{D}}$ was not determined).

\begin{table}[tbh]
\centering
\caption{\label{table:app:elconst:fcc}Elastic constants, bulk moduli, and Debye temperatures for metastable fcc structures with equilibrium lattice parameter $c_{\text{fcc}}$ (cubic axes) obtained with PW92. Ferromagnetic order is indicated (fm).}
\begin{ruledtabular}
\begin{tabular}{l*{7}{r}}
element & \multicolumn{1}{c}{$c_{\text{fcc}}$} & \multicolumn{1}{c}{$B$} & \multicolumn{1}{c}{$c_{11}$} & \multicolumn{1}{c}{$c_{12}$} & \multicolumn{1}{c}{$c_{44}$} & \multicolumn{1}{c}{$\theta^{\text{e}}_{\text{D}}$} & \multicolumn{1}{c}{$\theta^{\text{p}}_{\text{D}}$}\\
& \multicolumn{1}{c}{$(\textrm{\AA})$} & (GPa) & (GPa) & (GPa) & (GPa) & (K) & (K) \\
\hline
Sc & 4.477 & 59  & 81 & 47 & 42 & 348 & 314 \\ 
Ti & 4.006 & 123 & 147 & 109 & 60 & 367 & 346 \\ 
Co (fm) & 3.429 & 273 & 382 & 216 & 194 & 578 &590  \\ 
Y & 4.912 & 46 & 67 & 35 & 40 & 251 & 257 \\ 
Zr & 4.429 & 103 & 124 & 91 & 46 & 251 & 224 \\ 
Tc & 3.828 & 338 & 495 & 261 & 213 & 520 & 519 \\ 
Ru & 3.762 & 360 & 560 & 261 & 304 & 592 & 621 \\ 
Hf & 4.363 & 119 & 158 & 95 & 70 & 228 & 202 \\
Re & 3.875 & 395 & 610 & 286 & 252 & 427 & 424 \\ 
Os & 3.817 & 432 & 685 & 291 & 367 & 486 & 510 \\ 
\end{tabular}
\end{ruledtabular}
\end{table}

%\bibliography{/home/stephan/bibliography/mybibliography}


\end{document}